\documentclass[12pt,cites,graphicx]{article}
\usepackage{epsfig}

\title{Analysis of an algebraic model for the chromophore vibrations of
CF$_3$CHFI}

\author{C. Jung $^a$, C. Mejia-Monasterio$^a$\footnote{present address:
Center for Nonlinear and Complex Systems, 
Universita dell' Insubria, Via Vallegio 11, 
22100 Como, Italy}, 
and H. S. Taylor$^{b}$\\[3mm]
$^a$ Centro de Ciencias Fisicas, UNAM \\
Av. Universidad, 62251 Cuernavaca, Mexico  \\[1mm]
$^b$ Department of Chemistry \\
University of Southern California \\
Los Angeles, California 90089 }

\begin{document}

\maketitle

\renewcommand{\thefootnote}{\fnsymbol{footnote}}

\begin{center} {\Large Abstract} \end{center}

We extract the dynamics implicit in an algebraic fitted model Hamiltonian
for the hydrogen chromophore's vibrational motion in the molecule $CF_3CHFI$.
The original model has 4 degrees of freedom, three positions and one
representing interbond couplings. A conserved polyad
allows the reduction to 3 degrees of freedom.
For most quantum states we
can identify the underlying motion that when quantized gives the said
state. Most of the classifications, identifications and assignments
are done
by visual inspection of the already available wave function semiclassically
transformed from the number representation to a representation on the
reduced dimension toroidal configuration space
corresponding to the classical action and angle variables. The
concentration of the wave function density to lower dimensional subsets
centered on idealized simple lower dimensional organizing structures and the
behavior of the phase along such organizing centers already reveals
the atomic motion. Extremely little computational work is needed.

\newpage

\section{Introduction}
\label{intro}
In recent years we have developed methods to investigate
algebraic models ( spectroscopic Hamiltonians ) for the vibrations
of molecules \cite{DCO,N2O,CHBrClF,C2H2a,C2H2b}.
The Hamiltonians reproduce and encode by their construction
the experimental data
\cite{expDCO,expN2O,expCHBrClF1,expCHBrClF2,expCHBrClF3,expC2H21,expC2H22}. 
In the mentioned examples the system is reducible to two degrees of freedom.
Therefore it is interesting to investigate systems where after all possible 
reductions 3 degrees of freedom remain in order to demonstrate that our methods
also work well for higher dimensional systems. In this paper we use as example
of demonstration the molecule CF$_3$CHFI for which in Ref. \cite{expCF}
an algebraic model has been presented for the motion of the H atom 
( chromophore ). This model has 4 degrees of freedom and it has one conserved 
polyad which can be used to reduce it to 3 degrees of freedom. 

For the already known wave function ( they are needed to construct the
Hamiltonian ) transformed from the number representation so as to be able 
to be plotted in the toroidal configuration space
of action/angle variables \cite{smc} we show that we can visually sort
most of the states into ladders of states with similar topology. Each
ladder has a relatively simple spectrum. Complexity arises from their
interleaving. We also can recognize underlying classical lower dimensional 
organizing structures by the fact that for these inherently complex
functions the density ( magnitude squared ) is concentrated around them 
and the phase
has simple behavior near them. The position of these structures in angle
space reveals the nature of the resonance interaction causing the
topology. It also allows the
reconstruction, often by analytic use of the transformation back to 
generalized coordinates, which are near displacement coordinates, 
of the motion of the atoms which underlie this
particular quantum state. By counting nodes in plots of the wave function
density and phase advances in corresponding plots of the phase
quantum excitation numbers can be obtained.
 In many cases, along with the polyad quantum number itself, we thereby obtain 
a complete set of quantum numbers for a state even though the
corresponding classical motion is nonintegrable. These classification
numbers can be interpreted as quasi conserved quantities for this
particular state or for the ladder of states based on the same
dynamic organizational element. Note there are usually several different
types of organizing structures and hence ladders in the same energy region. 

Why are the wave functions that are so visually complex as to be
unsortable in displacement coordinate space simple enough to be sorted 
in our semiclassical representation ? The answer is fivefold. 
First, when a spectroscopic Hamiltonian is used and 
as energy changes frequencies can move into resonance, that is rational 
ratio, a constant of the motion called the polyad number comes into 
existence ( see section 2 ). It can be used to reduce the dimension
of the configuration space when the action angle variables are used, which come
out in taking the semiclassical limit of the spectroscopic Hamiltonian.
 Dimension reduction is always a great simplification.
Second, the resonant form of the fitted spectroscopic Hamiltonian can be 
justified formally by using Van Vleck or canonical perturbation theory  
starting with a given potential surface \cite{expCF,Joyeux,Ezra}
in a way that takes
in only the dominant dynamics. The terms that are left out for the most 
part are multiple small perturbations that in turn cause multiple small
distortions to the wave function. This makes them harder to recognize but
not fundamentally different. An excellent example of this is seen in
\cite{expCHBrClF2}, section 3 and quite pictorially in \cite{jfs} 
where figures 4 and 9 show the wave functions for the same
states but computed from the full and the spectroscopic Hamiltonian
respectively. The simplicity of the latter is enough to illustrate
our point. The third reason for the relative wave function simplicity
is that as the bound wave functions lie on a toroidal configuration
space their parts must interfere constructively as any angle is
advanced by $2 \pi$. This coherence results in simpler but not always
simple wave functions. The fourth reason is in the freedom to choose a
particular type of reduced dimension canonical variables. These variables
are designed with guidance from the polyad and the angle dependence of
the resonance to have zero velocity when the phase locking caused by 
the resonance sets in. The wave function in reduced configuration 
space will then in turn further collapse onto the environs of ( in the sense
of small oscillations about ) even lower dimensional surfaces or lines.
We defer explaining and exhibiting this point till section 5 which
introduces needed notation. The fifth reason for simplicity is that, 
as alluded to above, the wave function is complex and can then be 
plotted separately in density and phase. Each part helps in visually
sorting of wave function topologies and therefore dynamics and each
yields the values of different quantum numbers.

As organizational elements we do not necessarily use actual
classical periodic orbits. In many cases it is simpler to study
the general ( average ) classical flow and to relate to it some
idealized guiding center and to use this idealized motion as the
skeletal element on which the quantum density is concentrated.
We can imagine that this idealized structure contains the essence
of a large number of periodic orbits performing on average
a similar motion. The quantum dynamics is not able to resolve all the
fine details of the classical motion, nor the infinity of long
periodic orbits, nor their enormous number of bifurcations. 
Its limited resolution only recognizes the average of the classical 
dynamics over cells of the size of its resolution.
Therefore we interpret the quantum state as based on the motion of the
atoms which corresponds to the idealized structural element.
In many cases one state of the group of similarly organized states can be
interpreted as the quantum ground state on this element and the other
ones are various longitudinal and transverse 
excitations of the
basic structure, hence the ladders. In some cases only
sections of a ladder are found.
 
Further in this paper, we see how the higher number and smaller spacing
between observed states of a
system with more degrees of freedom makes the whole investigation
more complicated and in particular how the higher density
causes more mixing of structures. This increase of structural
mixing might be interpreted as one way in which effects of classical 
chaos enter the quantum behavior and results in a group of states
being unassignable even in the {\bf semiclassical angle representation} 
that brought simplicity in appearance to the density and phases of the
assignable states.

In section 2 we discuss the spectroscopic Hamiltonian, section 3
presents the semiclassical representation of the wave functions on the
toroidal configuration space and section 4 contains the dimensional reduction.
Section 5 explains the basic ideas how to extract the assignment of
a wave function from its plot on the reduced configuration torus.
Section 6 explains how from information about the dynamics in the
reduced system we go back to reconstruct information about the
original system.
Section 7 contains the classification of states and section 8 gives
final remarks.

\section{The model}

For the description of the chromophore dynamics of the molecule
CF$_3$CHFI we use the model set up in \cite{expCF}. It is based on
4 degrees of freedom as required by the nature of the observed overtone 
spectroscopy. Roughly they describe the three degrees of freedom
of the H atom. The index s stands for the stretch, the index a labels
the bend in the HCCF$_{(4)}$ plane and the index b labels
the bend perpendicular to this plane, the index f stands for the 
transpositioned C-F stretch. For the
exact atomic motion belonging to these 4 degrees of freedom see Fig.2
in \cite{expCF}.
The algebraic Hamiltonian fitted to experiment and to calculations
on a fitted potential surface has a natural decomposition
\begin{equation}
H_q = H_0 + W
\end{equation}
into a diagonal part 
\begin{equation}
H_0 = \sum_j \omega_j a^{\dagger}_j a_j + \sum_{ j \le m} x_{j,m}
a^{\dagger}_j a_j a^{\dagger}_m a_m
\end{equation}
(here the indices j, m run over the 4 degrees of freedom a, b, f, s) 
and an interaction part
\begin{equation}
W = \sum_{j,m} k_{s,j,m} [a_s a^{\dagger}_j a^{\dagger}_m +
a^{\dagger}_s a_j a_m]/(2 \sqrt{2}) +
\end{equation}
\begin{equation}
\gamma [a_a a_a a^{\dagger}_b a^{\dagger}_b +
a^{\dagger}_a a^{\dagger}_a a_b a_b]/2 +
\end{equation}
\begin{equation}
\delta [a_a a_a a^{\dagger}_f a^{\dagger}_f +
a^{\dagger}_a a^{\dagger}_a a_f a_f ]/2 +
\end{equation}
\begin{equation}
\epsilon [a_b a_b a^{\dagger}_f a^{\dagger}_f +
a^{\dagger}_b a^{\dagger}_b a_f a_f]/2
\end{equation}
where $a_j$ and $a^\dagger_j$ are the usual harmonic destruction
and creation operators of degree of freedom j. In Eq.3 the indices j and m
run over the 3 degrees of freedom a, b, f. For the coefficients
we use the ones given in column 6 in Table I in \cite{expCF} where
we see that roughly $\omega_s \approx 2 \omega_a \approx 2 \omega_b
\approx 2 \omega_f$ giving rise to the Fermi resonances anticipated by the
terms of Eq.3 and the Darling-Dennison resonances anticipated by the terms of 
Eqs.4-6. 
Eq.2 has first a ``harmonic oscillator'' term and a second ``anharmonic'' 
Dunham term.

The implied unperturbed basis of the space upon which $H_q$ operates is
the number basis
\begin{equation}
| \vec{n}\rangle = |n_s, n_f, n_a, n_b\rangle = |n_s\rangle |n_f\rangle |n_a\rangle |n_b\rangle
\end{equation}
$n_j$ are the quantum numbers or degree of excitation of the harmonic
mode $j$ and are eigenvalues of the number operator
$ \hat{n}_j = a_j^\dagger a_j $. $|n_j\rangle$ is its pure oscillator mode eigenfunction.
$H_0$ is diagonal on this basis. The unperturbed ( first term of $H_0$ )
energy of a basis function $|\vec{n}\rangle$ is 
$E^0_{\vec{n}} = \sum_j n_j \omega_j$. When frequencies are near resonant
as $\omega_s \approx 2 \omega_a \approx 2 \omega_b
\approx 2 \omega_f$ a polyad naturally arises with many consequences.
First we can write $E^0_{\vec{n}} = 2 $P$ \omega_b$ with the polyad $P$ defined as
\begin{equation}
P = n_s + ( n_f + n_a + n_b)/2
\end{equation}
with the corresponding operator given as $\hat{P}$ by replacing $ n_j
\rightarrow \hat{n}_j$ in Eq.8. Clearly all basis functions with $n_j$
giving the same value of $P$ are degenerate. If basis functions are sorted
by polyad, the Hamiltonian matrix which can be evaluated algebraically,
is expected by perturbation theory ( degenerate basis functions
interact most strongly ) to be nearly block diagonal suggesting that
$P$ is a near constant of motion. In fact, it is totally block
diagonal as $P$ is here an exact constant of the motion since $\hat{P}$ 
commutes with $H_q$ by construction \cite{expCF}, albeit not with the 
true Hamiltonian to which $H_q$ 
is, as said, to a high order of perturbation an approximation. When each 
block is diagonalized, each eigenfunction
\begin{equation}
| \Psi^P_j\rangle = \sum_{\vec{n} \in P} c^j_{\vec{n}} |\vec{n}\rangle
\end{equation}
in block $P$ has a good conserved quantum number $P$ which roughly measures 
via the equation for $E^0_{\vec{n}}$ the degree of excitation of the state 
in units of the lowest frequency. One can now study each polyad separately
( second consequence of $P$ ).
Here we will concentrate on the one 161 state
polyad $P=5$ which is the most complex one in the range of measured energy.

\section{Semiclassical representation}

Having the $\{c^j_{\vec{n}}\}$ now allows us to immediately obtain,
without any extra work, semiclassical eigenfunctions in action/angle
variables ( good to the lowest two orders in $\hbar$ ).
We do this because as we will see such eigenfunctions can be viewed due to
the existence of the polyad as effectively having one less dimension.
Hence the problem must now be transformed from the number representation
to a semiclassical action/angle representation. The drop in dimension is
characteristic of classical mechanics where a constant of the motion,
here the classical analogue of $P$, can by canonical transformation 
allow one to reduce the dimension of the phase space. Here the reduction
is from a four degree of freedom system with action/angle coordinates 
$\vec{I}$, $\vec{\phi}$ to a reduced system with specific value of $P$
and three degrees of freedom with action/angle coordinates $\vec{J}$,
$\vec{\psi}$. This of course will accomplish the first promised
simplification, that is dimension reduction. 

To make a transition to the corresponding classical Hamiltonian
we bring the operators in $H_q$ first into symmetrical order, 
and then apply the semiclassical substitution rules
\begin{equation}
a_j \rightarrow I^{1/2}_j \exp(-i \phi_j),
a^{\dagger}_j \rightarrow I^{1/2}_j \exp(i \phi_j) 
\end{equation}
where $I_j$ and $\phi_j$ are the classical action and angle
variables. The resulting classical Hamiltonian is using Eqs.2-6
\begin{equation}
H_0 = \sum_j (I_j -1/2) \omega_j + 
\sum_{j \le m} x_{j,m} (I_j - 1/2)(I_m - 1/2)
\end{equation}
\begin{equation}
W = \sum_{j,m} k_{s,j,m} I_s^{1/2} I_j^{1/2} I_m^{1/2} 
\cos(\phi_s - \phi_j - \phi_m) / \sqrt{2} +
\end{equation}
\begin{equation}
\gamma I_a I_b \cos( 2 \phi_a - 2 \phi_b) +
\end{equation}
\begin{equation}
\delta I_a I_f \cos( 2 \phi_a - 2 \phi_f) +
\end{equation}
\begin{equation}
\epsilon I_b I_f \cos( 2 \phi_b - 2 \phi_f ) 
\end{equation}
It has the conserved quantity 
\begin{equation}
K = I_s + [I_a + I_b + I_f]/2
\end{equation}
 Because of the symmetric ordering required for
the transition to classical dynamics the
zero point between the eigenvalues of $P$ and the corresponding value
of $K$ is shifted by 5/4.

The reader, as an aside, might wish to contemplate how easily this
transition was taken. For the more familiar $H=T+V$ Hamiltonian form
one must solve the Hamilton-Jacobi equation to get such variables; indeed
a difficult process even if the system is completely integrable.
The simplicity is due to the fact that, as said, this should be
rigorously done starting with a potential surface and using perturbation
theory. A price is paid for circumventing perturbation theory and using
a fitted spectroscopic Hamiltonian. This price is that the relation
of the $a_j^\dagger$ and $a_j$ to the generalized momentum and
position variables $p_j$ and $q_j$ is rigorously unknown. We will return
to this issue later when going back from reduced space to displacement
coordinates; our so called lift process.

If one now quantizes any single harmonic oscillator term in Eqs.11-15
using Schroedinger correspondence 
$ I_j \rightarrow - i \partial / \partial \phi_j$, $\phi_j \rightarrow
\phi_j$ it is clear that, since the angles define a now toroidal 
configuration space,
\begin{equation}
|n_j\rangle \rightarrow \exp(i n_j \phi_j), |\vec{n}\rangle \rightarrow
\exp ( i \vec{n} \cdot \vec{\phi})
\end{equation}
with $n_j$ corresponding to $I_j- 1/2$. From Eq.9 we get
\begin{equation}
\Psi_j^P \rightarrow \Psi_j^P (\vec{\phi}) = \sum_{\vec{n} \in P}
c_{\vec{n}}^j \exp(i \vec{n} \cdot \vec{\phi})
\end{equation}
That is, the semiclassical wave functions are already known and
given by Eq.18.

\section{Dimension reduction}

Dimension reduction can now be carried out by noting that, using 
the polyad of Eq.8 to eliminate the fast degree of freedom s, 
the j-th eigenstate can be written
\begin{equation}
\Psi_j^P(\vec{\phi}) = \exp (iP\phi_s) \sum_{n_a,n_b,n_f} c^j_{P,n_a,
n_b,n_f} \prod_{k=a,b,f} \exp ( i n_k (\phi_k - \phi_s /2))
\end{equation}
\begin{equation}
= \exp(i P \phi_s) \sum_{n_a,n_b,n_f} c^j_{P,n_a,n_b,n_f}
\prod_{k=a,b,f} \exp(i n_k \psi_k)
\end{equation}
\begin{equation}
= \exp( i P \phi_s) \chi_j^P (\psi_a, \psi_b,\psi_f)
\end{equation}

This shows that the eigenfunctions in a given polyad $P$ have a
common phase factor dependence on $\phi_s$ and really only depend
on $\chi_j^P$ which is a function of the three angles
\begin{equation}
\psi_j = \phi_j - \phi_s /2
\end{equation}
for $j=a, b, f$. These three angle make up a three dimensional
toroidal configuration space $T^3$ upon which $\chi_k^P$ is
situated. In fact $\chi_k^P$ is just the Fourier decomposition of
the reduced dimension wave function on the configuration torus.
More rigorously the wave function $\chi_j^P$ defined by Eqs.19- 21
actually is the semiclassical wave function after the classical
constant of motion $K$ is used to reduce dimension by a canonical
transformation. Using the angle transformation of Eq.22 suggested
above by the existence of $P$ plus the trivial transformation
\begin{equation}
\theta = \phi_s
\end{equation}
the canonical transformation permits that the new actions are $K$,
given by Eq.16 and
\begin{equation}
J_j = I_j
\end{equation}
again for $j=a,b,f$. $\theta$ is the cyclic angle.
Starting with equations 11 to 15 and using equations 16, 22-24
a new Hamiltonian in the six canonical variables $\vec{J}$, 
$\vec{\psi}$ albeit parametric in $P$ or $K$ can now be written so directly
that to save space we omit it. The torus $T^3$ is the 
configuration space. Using the Schroedinger quantization for the new
variables is only semiclassical, i.e. is correct in the two lowest
orders in $\hbar$. Its use gives the semiclassical expression
\begin{equation}
\chi^P_j(\psi_a, \psi_b, \psi_f) = \sum_{n_a, n_b, n_f}
c^j_{P,n_f, n_a, n_b} \exp(i(n_a \psi_a+ n_b \psi_b + n_f \psi_f))
\end{equation}
for the eigenfunctions on the reduced configuration space.

In the new coordinates the basis function with polyad number $P$ 
and three indices $n_a, n_b, n_f$ corresponds to the product of 
three plane waves or rotors as
\begin{equation}
\prod_{k=a,b,f} \exp(i n_k \psi_k)
\end{equation}
Each single plane wave factor loops the $T^3$ in the positive direction.
The rotor is that into which the single mode harmonic oscillator
has been transformed. Since Eq. 25 is a manifestly complex function
it can be written as
\begin{equation}
\chi^P_j ( \vec{\psi}) = | \chi^P_j(\vec{\psi})| e^{i \Phi^P_j(\vec{\psi})}
\end{equation}
where the phase function $\Phi^P_j(\vec{\psi})$ is given as
\begin{equation}
\Phi^P_j = \arctan( Im \chi^P_j / Re \chi^P_j)
\end{equation}
For the basis function of Eq.26 it is well to note for future
reference that the density is constant and the phase is
\begin{equation}
\Phi_{\vec{n}} = n_f \psi_f + n_a \psi_a + n_b\psi_b
\end{equation}

Symmetry properties of the system in the toroidal configuration space will be
used in the following. The Hamiltonian is invariant under a simultaneous 
shift of all angles $\psi_j$ by $\pi$.
Therefore the reduced configuration torus $T^3$
covers the original configuration space twice and all structures show 
up in double, even though 
they really exist only once. In addition the system
is invariant under a shift of any angle $\psi_j$ by $2 \pi$ and it is
invariant under a reflection where all angles go over into their
negatives.

\section{Methods to evaluate wave function plots}

It is seen that the simple basis functions have constant density and
 any eigenstate dominated
by a single basis function has a density without sharp localization.
Resonances as for example $2 \omega_a \approx \omega_s$ will be seen 
to cause localization about a line $\psi_a = constant$. This follows as
$\psi_a = \phi_a - \phi_s /2 = constant$ when differentiated with respect
to time gives the frequency relation. Hence it is seen that resonances
are associated with localization and the fact that $d \psi_j/ dt =0$.
It can be claimed that by using angle coordinates that slow to zero velocity
at resonance we here assure that the wave function will ``collapse'' onto
and about a lower dimensional subspace called the organization center.
This in reverse gives a way to recognize the influence of resonances,
namely localization of the wave function on the configuration torus. 
Each linearly
independent locking of angles therefore reduces by 1 the dimension of the
subset of configuration space around which the wave function is concentrated.

The fact that the wave function is on $T^3$, a three dimensional torus,
presents graphical and visual challenges when attempting to sort wave
functions by inspection. To meet these challenges we employ the fact 
that $T^3$ is 1:1 with a cube with sides of angle range $2 \pi$ and
identified opposite boundary points. That is a point on one surface of
the cube is the same as a point on the direct opposite side of the cube.
The density and phase are plotted in such cubes. The density plots now
over the interior of the cube are organized over the whole cube or about
interior planes or lines or points if zero, one, two or three independent
resonant couplings respectively are active. 
Over (or ``along'' in the case of a line) the organizing
structure a smooth density with no sharp localization should appear.

Nodal structures will be visible and countable in directions
perpendicular to the organization element and clearly will be
associated with a localized direction. The count of such transversal
nodes supplies for each direction of localization a transverse quantum
number $t$, that replaces an original mode quantum number $n$ which
has been destroyed by the resonant interaction. The wave function in all
the localized directions can be considered as qualitatively similar
( a continuous deformation ) of an oscillator state of the corresponding
dimension. By the superposition of the running wave 
basis functions with appropriate
weights and signs locally around the organization center the nodal
pattern of an oscillator is reproduced. The transverse quantum numbers
$t_k$ are the corresponding oscillator excitation numbers.

The phase functions $\Phi$ according to Eq.28 are smooth and close to 
a plane wave in the neighborhood of the organization center. They tend 
to have jumps by $\pi$ and singularities away from the center. Therefore
phase advances along fundamental cycles ( loops on the torus or lines that
connect opposite boundary points on the cube ) of the organization center
are well defined and are necessarily integer multiples of $2 \pi$.
They provide longitudinal quantum numbers $l_k$ which also replace the
interacting mode quantum numbers. 

Some states present what at first look appears to be a familiar but
unexpected phase pattern in that the phase picture is too simple.
Cases will appear when the density is localized but the phase picture 
has the globally striped flag appearance of the nearly noninteracting
mode case. The strangeness disappears if ideas from nonlinear
dynamics are employed. Nonlinear dynamics tells us that at low $P$ most 
of the motion takes place on what is called the primary zone of phase
space. And the now interacting states are quantization
of tori in this region. 
This zone has no central organizing structure other than the whole
configuration space itself. It does contain quasiperiodic orbits 
corresponding to the uncoupled modes. 

As $P$ changes coupling of a 
subset or even all modes can occur by at least two mechanisms.
First is one where the existing periodic orbits bifurcate and 
create new secondary zones based on newly created organizing structures
( periodic orbits, 2 tori, 3 tori e.t.c. ). Also new secondary zones 
may be created seemingly spontaneously due to what are called 
saddle-center bifurcations. In these secondary cases the motions will have
changed and states based on tori or near tori ( think about the
resolution of $\hbar$ ) will have density and phase diagrams as described
above with $(P,t_k,l_k)$ quantum numbers.

The second mechanism of coupling which will manifest itself in density
localization is by continuous deformation; a process visualized by 
mentally drawing the primary configuration on rubber sheets and
stretching and pulling the sheets in various directions till the
localized form is achieved. Under such continuous deformation the 
phase function $\Phi$ of the continuous,
single-valued complex wave function $\Psi(\vec{\psi})$ always must
change by an integer multiple of $2 \pi$ as its argument $\vec{\psi}$
goes around any closed loop $\gamma$ in configuration space.
In our case prior to deformation $\Phi = \vec{n} \cdot \vec{\psi}$ and
that integer is $n_j$ if the loop is in $\psi_j$ ( this loop will
be called $\gamma_j$ ) i.e. as
$ \psi_j \rightarrow \psi_j + 2 \pi$, $\Phi \rightarrow \Phi + n_j 2 \pi$.
More generally for any $\Phi$ this can be stated as the loop integral 
relation
\begin{equation}
m_j (\lambda)= \frac{1}{2 \pi} \int_{\gamma_j} d \Phi (\vec{\psi};\lambda)
\end{equation}
where $m_j$ is an integer and $\lambda$ is the deformation parameter.
Now under a continuous deformation it is clear that $m_j$ or in our case
$n_j$ must remain constant as a continuous integer valued function.
In our case this $n_j$ can be read off from the phase diagram as equal 
to the number of revolutions of the phase along the loop $\gamma_j$
achieved as $\psi_j \rightarrow \psi_j + 2 \pi$. For the noninteracting 
modes the afore mentioned quantum number $l$ is just the $n$.

From the point of view of basis state mixing even though the phase is
behaving as if a single basis function is totally dominant, significant
state mixing can occur. This can and will occur in proportion to the
amount of localization of the density. Weak localization tends to occur
when one basis function is dominant, the one with the same $\vec{n}$.

The $\vec{n}$ assignment holds till a small change in $\lambda$ causes
an abrupt change in $\Phi$, that is when singularities occur. At this
point continuous deformation will stop. Now since the wave function must 
be continuous, singularities in phase can only occur where the density
and hence the wave function is zero. For us this occurs at the nodes and
perhaps the outer edges of the localized oscillator. In short when 
nodes occur one can no longer use this simple full set of $\vec{n}$
quantum numbers. One can still count phase revolutions along loops inside
an organized region ( parallel to the organizing center ) to get the
$l$ quantum numbers. Great care should be exercised in deciding if for a
localized density the phase is a continuation of a plane wave or not. To
do this $\Phi(\vec{\psi})$ must be a smooth function in the 
neighborhood of the loop along which one is counting the phase revolutions.

\section{Ladders and motions in displacement coordinates}

Once the wave functions in their semiclassical reduced dimension
representation are diagrammatically presented and quantum numbers assigned 
as described in the previous section, as will be seen in the next
section, most often states can be visually sorted and ordered into
ladders of states with a given topology. Each ladder tends to lie in 
the Husimi sense in one of the phase space resonance zones discussed 
previously. The simplicity of sorting is due to the fact that as said 
an idealized classical organizing element underlies each ladder.
This structure can easily be written as a configuration space relation 
between the $\vec{\psi}$ variables. The structure will allow us to 
immediately determine what type of resonance interaction is creating this
structure. An approximate procedure that transforms the variables
$\vec{\psi}$ back to displacement space reveals the atomic motion
( modes of motion ) whose quantization yields the ladders of states. This
process we call ``lifting'' and refer to it as the ``lift''.
We now describe the lift in a way that starts with the least
approximate and most difficult to implement algorithm and proceeds to
simpler to implement ones that still retain the qualitative spirit of
the motions. 

To determine the key interactions we only illustrate the procedure with
a simple example. Assume that the organization structure is a line
( or fiber ) along $\psi_a = 0$. Since $\psi_a = \phi_a - \phi_s /2$
then $ ~\dot{\phi}_a - ~\dot{\phi}_s /2 \approx \omega_a - \omega_s /2 
\approx 0$ and the $2 \omega_a = \omega_s$ phase locking relation 
speaks to us of a 2:1 Fermi resonance between the bending mode a and
the stretch mode s. Other cases are similarly treated.

Now for interpretation of the motion a trajectory 
 in the $(\vec{J},\vec{\psi})$ phase space need be found which
 when projected into configuration space should represent the
organizing element. In reverse the knowledge of the organizing element
is helpful in finding an appropriate trajectory. To start this 
search estimates of the actions will be useful. For uncoupled modes the
assigned $n_j$ quantum numbers using $I_j = n_j + 1/2$ suffices. To obtain
estimates of the initial actions the quantum mechanical average of the
$J_j$ can be used and trivially computed from the known wave functions as
\begin{equation}
\langle J_j \rangle = \langle \Psi | - i \frac{\partial}{\partial \psi_j} | \Psi\rangle =
\sum |c_{\vec{n}}|^2 n_j
\end{equation}

The ladder state that is most localized should be used for this and
to check an estimate the variance can be used. With this we now have
initial conditions ($\langle J_j \rangle$ and a value of $\psi_j$ on the
organizing structure) with which to use Hamilton's equations to get 
$\psi_j(t)$, $J_j(t)$. Then first we reconstruct the cyclic
angle $\theta$ by integrating 
\begin{equation}
\theta(t)= \theta(0) + \int_0^t ds \frac{\partial H}{\partial K}
(\psi_j(s), J_j(s))
\end{equation}
Next we undo the canonical transformation to get the corresponding
$\phi_j(t)$ and $I_j(t)$. Then we assume some idealized harmonic
model for the elementary degrees of freedom and form the displacement
$q_j(t)$ and its conjugate momentum $p_j(t)$ as
\begin{equation}
q_j(t) = \sqrt{2 I_j(t)} \cos(\phi_j(t)),
p_j(t) = \sqrt{2 I_j(t)} \sin(\phi_j(t))
\end{equation}
 The functions $q_j(t)$ will be interpreted as
representatives of the atomic motion.

Diagrams in which the $q_j$ are taken as coordinates with t as 
parameter give the more conventional picture of the atomic motions.
In the simplified methods the average values of $I_j$ which 
giving the scale of the $q_j$ motion will replace $I(t)$ in Eq.33.
These methods are simple because again Eq.33 is used but
the time dependence and therefore the computation is suppressed and the
relations gotten between the $q_j$ from the organizing structure
equation that relates $\phi_j$ along with the use of Eqs.23, 24.

In preparation for section 7 let us discuss a few very
simple cases. First there is the possibility that the density is
distributed rather homogeneously over the whole $T^3$ and the
phase function comes close to a global plane wave with 3 component
wave vector $(n_a, n_b, n_f)$. This means no locking at all,
all modes run freely with their individual frequency and the
excitation numbers of the various modes are $n_j$ read off
from the wave vector for $j=a,b,f$ (i.e. count the revolutions
as one traverses the phase diagram in the j'th direction)
and $n_s = P -[n_a + n_b + n_f]/2$.
If the density is concentrated in a plane, e.g. the plane
$\psi_a=constant$ then the angle $\psi_a$ does not move in the 
long run. It only may oscillate around the value of the constant.
From the canonical transformation of Eq.22 we see that this means
$\phi_a - \phi_s/2 = constant$ for the angles of the original
degrees of freedom.
Therefore we have frequency and phase locking between the modes
a and s. The other modes b and f run freely and their excitation is 
according to the wave vector of the phase function in the plane
of high density. Analogous considerations hold for other planes
of concentration of the density.
If the density is concentrated along a line, e.g. the line
$\psi_a = constant$ and $\psi_b = constant$ ( this is a fiber
running in f direction ) then we see from
Eq.22 that $\phi_a - \phi_s/2$ and $\phi_b - \phi_s/2$ are
kept close to constant values and this means locking between
a and s and simultaneously locking between b and s. Of course,
then also a and b are locked relatively to each other. The mode
f runs freely in this case and its excitation number follows from
the wave number of the phase function along the fiber of high
density. Analogous considerations will be applied for a concentration
of the density along other 1 dimensional lines in reduced
configuration space. 

\section{The motion behind individual quantum states}

For polyad 5 with 161 states, the given transformations and equations
lead to a three dimensional toroidal configuration space which here is
a cube with coordinates $\psi_a$, $\psi_b$ and $\psi_f$. Three dimensional
( perspective )
plots of the semiclassical state functions look too often like large
globes of ink and were not informative. Here we will resort to cuts of the
cube ( e.g. cut $\psi_a = \psi_b$) and plot density and phase in such cuts.

The question is now where to cut. Clearly a knowledge of the classical flow
was most helpful here. An analysis as follows is also useful. The cuts should
contain organizing elements about which the density is localized. The phase
should be simple over the regions of the plane having high density. 
Note the idea of an organizing
structure is that one has found the configuration part of the quasiregular
resonant zone of the coupled modes. The phase simplicity can be used as
a tool for finding organizing structures if they exist as simple phase
patterns are immediately apparent to the eye. Here we have used
both trial and error as well as dynamics guidance to find the cuts.

Dynamics guidance means for example that in CF$_3$CHFI a resonance with a
rather close frequency is $\omega_f \approx \omega_b$ and its associated
$k_{b,f}$ in the Hamiltonian is largest. This suggests that modes b
and f should be coupled in the plane $\psi_f = \psi_b + constant$ which
assures $\omega_f^{eff} = \omega_b^{eff}$. Note the constant will vary
as explained below in a dynamically meaningful way, i.e. it will be used
for state classification on a ladder of states and allow the ladder to be
subdivided into similar sub ladders.

The Fermi choice $\omega_s = 2 \omega_a$ along with $\omega_s = 2 \omega_b$
would then give by similar reasoning the organizing line $\psi_a = c_1$,
$\psi_b = c_2$. For some states several organizing structures could be used
and the same dynamics revealed. Our experience is to choose those
corresponding resonances which seem more important in $H$ and which
give longer ladders of states.

\subsection{Scheme for a large number of states throughout the polyad}

Starting at the bottom of the polyad we find 64 states that
definitely lie in the Darling-Dennison $\omega_f = \omega_b$ resonance
class in that they have densities localized about planes all
denotable as $\psi_f = \psi_b + constant$.
They also have simple phase plots in these planes.
At the lower end a few of them could also be organized by the
$\omega_s = 2 \omega_f$, $\omega_s = 2 \omega_b$ Fermi resonances,
i.e. about a line in the ``a'' direction; but they occur when the stretch
excitation is zero and this classification is not very physical.

Some of the lowest states could also be assigned for all modes by
quantum numbers $n_j$ and correspond to continuous deformations of basis
functions. Although these $n_j$ are useful in the lift for obtaining actions
these assignment misses the phase and frequency locking and the localization
of the density. In addition we looked very carefully for primary tori
in the classical phase space at the corresponding energy and did not
find any. Therefore EBK quantization cannot be applied. Hence classical
dynamics shows that the assignment by $n_j$ for $j=a,b,f$ is unphysical.

In what follows we shall first discuss motion in the planes
$\psi_f = \psi_b + const.$ This leads us to conclude that modes ``a''
and ``s'' are uncoupled and that the Darling-Dennison resonance couples
``f'' and ``b'' such that mode ``b'', a true mode of the Hydrogen atom
motion, is driven by the generic source mode f which represents
the effect on the Hydrogens motion of the motion of the rest of the molecule.
At this point in our analysis the quantum numbers $n_s$, $n_a$ and $l_{f+b}$
will first be assigned to each state as explained by the examples below.
$l_{f+b}$ represents the number of quanta in the DD coupling. Clearly
one more quantum number is needed for a full assignment and this
would be the transverse one representing the degree of excitation for
the localized motion transverse to the plane $\psi_f = \psi_b + const.$ 
In most prior papers in this series this quantum number, $t$, could be counted
as the number of nodes perpendicular to the organizing structure.
In our analysis of the CDBrClF \cite{cd} 
spectra we were unable to consistently
see these nodes clearly but could assign $t$ as a numerical excitation index
based on the distance of density maxima from the $\psi_f = \psi_b + const$
organizing plane. Here the situation was not as opaque as nodes could
be seen. What at first was confusing was that they appeared very
often to be organized about one of two parallel planes and at higher energy 
about third and fourth planes. 
At low energy this observation and the fact that
pairs of states with identical ($n_s$, $n_a$, $l_{f+b}$, $t$) assignments appeared,
led to the discovery of a dynamic nearly symmetric double well or double
``valley''. Two organizing planes at $\psi_f = \psi_b$ and 
$\psi_f = \psi_b + \pi$, each run along and bisected the transverse width
of one of the valleys. The $\psi_a = \psi_b$ was slightly deeper making
the $\psi_f \rightarrow \psi_f + \pi$ invariance be only approximate.

For some of the lower states, the energy in the ``mode f - mode b'' 
lock is low enough that these states are primarily localized in the 
$\psi_f = \psi_b$ or the $\psi_f = \psi_b + \pi$ well. 
As energy increases the states are roughly
speaking ``above the barrier'' of the ``double well'' with density on both 
sides and they fall into near symmetric ($+$) and antisymmetric ($-$) 
pairs (with the same $n_s$, $n_a$, $l_{a+b}$, $t$) reflecting the approximate 
invariance $\psi_b \rightarrow \psi_b + \pi$. Each member of these pairs 
has a slight density preference for one of the wells. 
The ladders with an even $l_a$ start with a $+$ state and ladders with an
odd $l_a$ start with a $-$ state. The lowest state in any ladder of
constant $l_a$ always has most of its density near the plane
$\psi_f = \psi_b$.

Clearly it can be said that roughly two ladders
exist. Each represents states in b-f lock but with motions
that are $\pi$ out of phase with each other. In Table I this classification
($\pm$) label appears. It is necessary for classification as the other
four quantum numbers label only pairs of states. It is also necessary to 
overcome the confusion that arose in trying to count the transverse
modes. This count now makes sense once the organizing plane is specified
and it is noted that within a pair the $t$ value in column 6 is common.
For higher states it is sometimes becomes difficult to count $t$. The reason
for this can be traced to the fact that a slice transverse to the organizing
plane should ideally reveal wave functions that represent oscillators.
The density would then be the highest in the two planes running parallel and 
on opposite sides of the organizing plane. Now consider that for higher states
the density is localized about both planes (albeit with a preference
for one plane). As excitation increases then the outer lobes of the transverse
oscillator will move toward the planes in between 
(recall all features appear by symmetry in doubles)
the two original planes
i.e. the planes $\psi_f = \psi_b \pm \pi/2$ on which the highest density
will now accumulate.
Hence it now make sense to count transverse nodes between these now bigger
intensities. We label these as $t'$. Sometimes both $t$ and $t'$ assignments
are possible. Figure 2c and 2d show the $t'$ assignment. To see it a 
$\psi_f = \psi_a + \pi$ plane that cuts the $\psi_f = \psi_b + \pi/8$ plane
is chosen.

As always the transverse quantum number indicates to which extent the coupled
motion goes out of exact phase lock. It is a measure of the width of the 
phase distribution.

A feature that further complicates and hides the true 
interpretation of the states is mixing due to 
the accidental degeneracies of two or three states.
The mixed final eigenstates that result are spotted by noting their near
degenerate energies and their lack of almost all of the above discussed
features. Trial and error demixing using various weights yields states with
the above features. The notation $s/s'$ or $s/s''/s''$ indicates in Table I
that this particular state is a demixing of $s+s'$ etc.

The main method used for assignment, at least initially, is simply to
inspect various 2D cuts of wave functions that hopefully exhibit
recognizable localizations, nodes and plane wave structures. This ``informed''
inspection and slicing of the configuration cube is aided by first
inspecting the lower levels on the ladders which exhibit simpler
behavior. Further insight is gained by running 
many long classical trajectories that show here that the motion,
that is the flow of the trajectories, is mainly parallel to the
$\psi_f = \psi_b$ plane and in addition is mainly in the 
direction of the space diagonal i.e. guided by the condition
$\psi_f=\psi_b=\psi_a$. This is not totally unexpected as 
$\omega_a\approx\omega_b\approx\omega_f$. This effect is reflected
in the observation that in the organizing planes 
$\psi_f = \psi_b$ and $\psi_f = \psi_b + \pi$
the wave function tend to have fibers of density running along the diagonal.
In fact a diagonal classification could have served many states as the
organizing structure and assignment might have been made relative to it.
We chose the parallel planes as the organizing structures as breaks in this
diagonal fibration often made assignment less than clear.

The ``double well'' observation was first made 
by constructing an ``accessibility
diagram''. This is a plot of regions of configuration space, i.e. the
values of $\psi_a$, $\psi_b$ and $\psi_f$ that satisfy 
$E=H(J_a,J_b,J_f,\psi_a,\psi_b,\psi_f;P)$, at each $E$, for given $P$ and any 
compatible actions. Around 11100 cm$^-1$ (well below the lowest quantum state
of Polyad 5 and near to the classical lower end of Polyad 5)
it was noted that only
two slabs of configuration space were accessible; these are our wells.
At the energy of the lowest quantum state all configuration space is 
accessible but the wells act as two attractive wells that cause quantum
localization above them.

To illustrate our ideas let us consider two states, state $s=7$ and
$s=43$ at the bottom and the middle of the ladder respectively.
Both states are organized about $\psi_f=\psi_b$ giving a DD 
$\omega_f=\omega_b$ expectation. State 7 is a low state with its density
running up the valley in the middle of the well as seen in Figures 1c and 1a.
Recall again
that due to the symmetry property discussed at the end of the section 4
all features come in double. Figure 1a looks down on the organizing plane
and figure 1c looks sideways. In Figure 2 the highly exited case discussed
above is exhibited. The reader here can without seeing it be assured that
the highest density is in the plane $\psi_f=\psi_b+\pi/8$.
Figures 2a and 2b look down on this plane of high 
density and figures 2c and 2d show a view from the side 
that shows transverse excitation.

With this we proceed to extract the quantum numbers $n_s$, $n_a$ and 
$l_{b+f}$ for the two example cases of states 7 and 43. Figures 1 and 2
show that if one stays on the high density plane and heads say in the a
direction the torus is looped during a $2 \pi$ change of $\psi_a$.
This means that the $\psi_a$ motion is free and that on the phase plot
following any path with $\psi_a \rightarrow \psi_a + 2 \pi$ the
free $\exp(i n_a \psi_a)$ factor in the wave function should go to
$ \exp(i n_a (\psi_a + 2 \pi))$. In Fig.1b or Fig.2b respectively,
as $\psi_a \rightarrow
\psi_a + 2 \pi$ and if we follow a line in the ``a'' direction one sees 
a phase advance of $4 \pi$ and $2 \pi$ respectively. We conclude that
$n_a = 2$ for state 7 ( Fig.1b) and $n_{a} = 1$ for state 43 ( Fig.2b).
Sometimes here $n_a$ will be usefully called $l_a$ to emphasize that
it is a longitudinal quantum number obtained by following the path of
highest density around the ``a'' loop.

The phase change accompanying a loop of the b direction in figures 1b and 2b
by the nature of the cut causes a looping on the $\psi_f = \psi_b + constant$
plane in the complete configuration space. This looping advances both $\psi_f$
and $\psi_b$ by $2 \pi$ and on this plane the wave function must change
as $ \sum c \exp(i \vec{n} \cdot \vec{\psi}) \rightarrow
\sum c \exp(i (\vec{n} \vec{\psi} + 2 \pi ( n_f + n_b)) $
The choice of the $\psi_f = \psi_b + constant$ plane required that
we see a simple plane wave factor $\exp(i(n_f \psi_f + n_b \psi_b))$
which means that all the important basis states with large mixing coefficients
must have this factor. Hence the wave function phase must advance as
we sweep $\psi_b$ in this organizing plane by $2 \pi (n_f + n_b)$.
We denote this by $l_{b+f}$, $l$ for longitudinal, i.e. motion along the
organizing center as opposed to transverse to it. This longitudinal
quantum number clearly tells us the total number of quanta in the Darling
Dennison lock.

For state 7 in figure 1b $l_{b+f} = 8$ and for state 43 in figure 2b
$l_{b+f}=7$. Using the polyad expression for the stretch mode we get
$n_s = P - (n_a + l_{b+f})/2 = 0$ for state 7 and $n_s = 1$ for state 43.
 Now a partial ( three out of four
needed quantum numbers ) assignment can be made as $(P,l_a = n_a, n_s)$
or equivalently $(l_{b+f},l_a=n_a,n_s)$. State 7 is now $(5,2,0)$ or $(8,2,0)$
respectively and state 43 is $(5,1,1)$ or $(7,1,1)$ respectively as recorded in
table I. In principle $n_f$ and $n_b$ no longer exist as good quantum
numbers except when as explained these numbers represent conserved
integer loop integrals as in the continuous deformation case for state
7 where both phase diagrams ( Figs 1b and 1d ) look as they represent 2D
plane waves. Figure 1d then tells us that $n_b =8$ and $n_f=0$. This is
useful only for obtaining $I_b = n_b + 1/2 = 17/2$ and $I_f = n_f + 1/2 = 1/2$
which will in turn be useful in the lift process.

For state 43 phase simplicity only happens in the organizing plane.
Fig. 2d shows no useful phase information. The plot shows jumps along lines
and ramp singularities where the phase value depends on the direction
of approach.

At this point and in a similar way a full 64 states for $P=5$ can be
assigned by $n_s$
and $n_a$. Actually an additional 38 states seem to resemble
this picture. The resemblance is only seen as trends from the systematic
following of the 64 other ones. For these, increasing but still smaller
effects of many other resonances are taking their toll and they
significantly degrade the assignable states into unassignable more heavily
mixed quantum analogues of classical chaos. The table I indicates
which states have less certain assignments.

Subladders of common values of $n_a$ and $n_s$ (or equivalently $l_a$ and
$l_{f+b}$) can now be made. To complete this part of the table demixing
must be used and the table indicates states which result from the process.
Each subladder now needs classification according to ($+$) and ($-$)
and assignment of $t$ and/or $t'$. 

By looking at the wave function amplitudes above the $\psi_f=\psi_b$
(plane 1) and $\psi_f=\psi_b + \pi$ (plane 2) wells for each pair of
states with the same $(n_s,n_b)$ the ($+$) and ($-$) assignment can
be made. It is seen that state 1 is concentrated above plane 1 and
state 2 about plane 2; the latter is a little higher in energy.
State 3 is about plane 1. The nearly degenerate states 4 and 5 need
demixing. Then (4-5) is concentrated about plane 2, and (4+5) is about
both planes. As described above higher excitation leads to states over both
wells but with a slight favoring of one well.

To see the transverse structure of the states we need plots of the
density in some appropriate plane transverse to the plane of high density.
In figure 1c we see that state 7 has $t=0$ and in Fig.2c we see that state
43 has $t=1$. The number $t'$ does not make sense for state 7 and the
number $t'=2$ for state 43 is rather unclear from Fig.2c. However the phase
plot in Fig.2d shows two lines of phase jumps near the lines
$\psi_a = \psi_b \pm \pi/2 $ and these lines of phase jumps indicate nodal
lines of the density. Since this is a somewhat indirect indication of the
value $t'=2$, in table I the number $t'=2$ for state 43 is set in brackets.

Even at this stage where other ladders discussed below are omitted spectral
complexity arises from the interlacing of energy levels of different
subladders. The roughly 100 states on the ladder which only demands the
DD-f/b lock for its construction gives a complex spectrum and nontransparent
energy spacings. In addition the individual ladders appear irregular
by the mixing of states. A sequence of states which are pure excitations
of a single organization element would form a regular ladder of energy levels
looking like the spectrum of one anharmonic oscillator.
However by the mixing the eigenstates are mixtures of various pure excitations
and thereby their energies are shifted and thereby the ladders appear
more irregular. 

The reader will note that some ladder states in table I are missing.
These states could not be found but their disappearance is confirmed
using the anharmonic models expected energy spacing. Formally these states
lie among the highly mixed ( chaotic ? ) states we discuss below. They
are mixings of more than two dynamically identifiable states and do not
show the near degeneracy that our demixing process identifies.
Parenthetically we note that the ultimate demixing to dynamically
recognizable states is the known inverse transform of the eigenstates
back to the dynamically uncoupled basis functions. The magnitude of the
eigenvector transformation elements would then have to be used for state
classification and as said at the very beginning, this is impossible to do
for the great majority of states.

The actions $I_j$ can be gotten for the modes s and a using
$I_j = n_j + 1/2$, in addition we have $I_f + I_b = l_{b+f} + 1$.
Then $I_b = J_b$ and $I_f = J_f$ is gotten from the state functions
using Eq.31. To observe approximate averages of the motion of the hydrogen
atom the $I_j$ can be used in Eq.32 for $\theta$ to enable us to get
$I(t)$ and $\phi(t)$ and then using Eq. 33 to draw qualitatively pictures
for the various pairs $(q_j,p_j)$. Actually a great deal can be said without
the integration for $\theta$. In $(q_s,q_{a/b/f})$ planes a roughly
rectangular region is seen whose sides are proportional to
$\sqrt{2 I_s}$ and $\sqrt{2 I_{a/b/f}}$. The hydrogen trajectories
motion interior to the rectangle will be quasiperiodic motion.
The exact trajectory depends on the initial choices of $\phi_s$ and
$\phi_{a/b/f}$ and is really not important. In the $(q_a,q_b)$ plane
organization structure leads to ellipses whose relative ranges
( excursion from zero displacement ) in the $(q_a, q_b)$ variables is
$\sqrt{I_a/I_b}$. The ellipses eccentricity depends on the organizing
structures phase shift, viz. $\psi_a = \psi_b + \delta$ $\rightarrow$
$\phi_a = \phi_b + \delta$. $\delta=0$ and $\pi$ give zero angular
momentum and something approaching a straight line motion while
$\delta = \pm \pi /2$ gives maximal angular momentum and elliptical
motion which becomes circular if $I_a = I_b$. The $t$ value is the out
of phase motion and causes the trajectory to change slowly its eccentricity.

\subsection{Scheme for the upper end of the polyad}

At the upper end of the polyad we take to start states 160 and 161.
A cut in $\psi_b =0$ ( or any other value of $\psi_b$ ) reveals as we see
in Fig.3 for state 160 what looks like two columns of density
localized around $\psi_a=\pi$ and $\psi_f=\pi$. In comparison state
161 ( not shown ) only has one column. A cut in the plane $\psi_b =
constant$ is called to see that the columns really exist. Figure 3 shows
that they do for state 160 and it also holds for state 161. Clearly for
both these states the line $\psi_a=\pi$, $\psi_f=\pi$ is the organizing
structure. Since it loops in $\psi_b$, mode b is decoupled and the phase
picture for both states shows ( figure 3 for state 160, state 161 not shown )
that as $\psi_b \rightarrow \psi_b + 2 \pi$ along the organizing line no
wave fronts are crossed and hence there is no phase advance. Hence
$l_b = n_b = 0$. For the upper end of the polyad also classical dynamics
shows a flow in b direction. As always the figures show all structures in
two copies which in reality are only one structure.

Now the $\psi_a = \psi_f = \pi$ organizing structure
implies that the phase locks are $\phi_a = \phi_s/2 $ and
$\phi_f = \phi_s /2 $. Time differentiation of these relations show
that the frequency locks are of 2:1 Fermi type with the stretch mode
locking with the bend mode a and the mode f. Clearly for the
organizing structure
$\psi_a = \psi_f$ then $\phi_a = \phi_f$ is a valid phase lock and
$\omega_a = \omega_f$ is a DD 1:1 frequency lock. Since all three types
of terms appear in the Hamiltonian we can be assured that all three
frequencies are in lock and that the quantum numbers $n_a$, $n_f$ and
$n_s$ no longer exist. They must be replaced by three quantum numbers
of the lock one of which can be $P$ and the other two can be taken as
the number of nodes seen in the $\psi_b=0$ cut along the antidiagonal as
$t_1=1$ and along the diagonal as $t_2=0$. For state 161 since one
column exists $t_1 = t_2 =0$.

If energy decreases from the top,
then the classical flow turns into the space diagonal
direction rather soon and accordingly only the 4 highest quantum states
show ( partly after demixing ) a clean fiber structure in b direction.
The fiber is always sharp in the a direction but broad in the f direction. This
demonstrates again an emerging  b-f mixing in the system.
The next 7 states still show some tendency towards fibers in b direction
( they are no longer really clean ),
but at the same time also show the beginning of fibers in the diagonal 
direction.
Their assignment as states organized along b fibers is less clear but still
they continue the trends seen in the 4 highest states ( partly after demixing ).
The corresponding assignments for the 11 highest states are compiled in table II.

An idea of the internal motion of the deuterium and the CF stretch can be
obtained from the actions, phase relations and the $t_1$ and $t_2$ values.
In all $(q_i,q_j)$ planes the actions again define the maximal displacement
$\sqrt{2 I_j}$ of each $q_j$ from its equilibrium. Clearly this range is
meant in the sense that an average is made over the values of the two
variables not represented in the plane. The plane $(q_b,q_j)$, $j=a,f,s$
will show rectangularly bound quasiperiodic motion. In the
$(q_s,q_{a/f})$ plane a U shaped region is swept out. Here $q_{a/f}$
reaches its extreme values as $q_s$ reaches its maximum displacement and
$q_s$ sweeps its range twice for each sweep of $q_{a/f}$. Increasing
$t_1$ and/or $t_2$ causes the lifted trajectory to oscillate in a tube
about this basic U lift. Of course $q_f$ and $q_a$ reach their maxima
( minima) in phase as the three modes s, a and f are phase locked.

\subsection{Patterns in the dense region of the polyad}

In the middle of the polyad there are approximately 35 states with very
complicated wave functions which we could not classify in any scheme.
It is possible
that further and closer analysis could reveal some systematics.
These states are dispersed among the states mentioned in subsection 7.1.
But interestingly a few simple states which do not
fall into any scheme discussed so far are dispersed in this region.
In this subsection we present a few such states.

Figure 4 shows the structure of state 82, parts a and b give density and
phase respectively in the plane $\psi_f = 0$ and parts c ans d show the
plane $\psi_b = \psi_a + 0.29 * 2 \pi$. Part a demonstrates some concentration
around $\psi_a = \pm \pi/2$ and at the same time around $\psi_a = \psi_b
\pm \pi/2$ with a transverse excitation $t=1$. Plots of the phases in
various planes show that planes $\psi_b = \psi_a + constant$ are the
only planes with a simple phase structure. Figure 4d shows it in the
plane of highest density. The phase function comes very close to a plane
wave with $l_{a+b} = 8$ and $l_f = 2$. Together with the above mentioned
transverse quantum number $t=1$ this gives the complete assignment of the
state. In Fig.4 notice that also the interaction between the degrees of
freedom f and s has an effect, it concentrates the density mainly around
$\psi_f = 0$ and $\psi_f = \pi$.
There are a few other examples ( states 93, 111, 114, 122 and 136 )
which show such a clean phase function in planes $\psi_b = \psi_a \pm
\pi/2$ but do not show a clean phase function in any other plane.
There are approximately 5 other states where the phase function is not
as clean as in figure 4d but comes close to a continuous deformation
of a plane wave only in a plane of this type. The motion of all these states
is a circular motion of the H atom, two orthogonal bends with relative
phase shift locked at $\pm \pi/2$.

Numerical results for state 80 are shown in figure 5, part a and b show
density and phase respectively in the plane $\psi_b = \psi_a + \psi_f +
\pi/2$ which is the plane of high density and thereby the organizing
structure. Part c and d show density and phase respectively in the
transverse plane $\psi_f = 0$. In the plane of high density the phase
function comes close to a plane wave with $l_{a+b} = 6$ and $l_{f+b} = 5$.
This is the only plane where the phase function is a continuous deformation
of a plane wave. This case is something new in so far as there is a plane
with a simple phase function and at the same time not a single $n_j$ has
a definite value. There is a locking involving all modes without
implying a locking in any pair of modes.
A few other states ( 75, 83, 87, 118 ) fall into the same scheme, i.e.
they have simple phase functions only in a plane of the form
$\psi_b = \psi_a + \psi_f + constant$.
The implications for the motion of the atoms are as follows. In the nonreduced
variables the locking condition is
$ \phi_b + \phi_s /2 - \phi_a - \phi_f = \pm \pi/2$. This can be interpreted
as a locking of the beat frequency between a pair of degrees of freedom to
the beat frequency of the remaining pair of degrees of freedom. We have several
possibilities to group the 4 modes into two pairs.

\section{Conclusions}

We have shown how our previously developed methods can also be applied
to systems with 3 degrees of freedom after reduction. For 75 percent 
of the states the strategy works successfully. We investigate
the wave functions constructed on a toroidal configuration space.
We investigate on which subsets the density is concentrated and
evaluate the phase function on this subset of high density. This
allows us to obtain a set of excitation numbers for the states; in
the best cases we obtain a complete set of excitation numbers which provide
a complete assignment of the state. The excitation numbers can be
interpreted as quasi conserved quantities for this particular state.

The success of the method is based on 3 properties of our strategy
which more traditional approaches starting from potential surfaces do
not have. First, we start from an algebraic Hamiltonian which naturally
decomposes into an unperturbed $H_0$ and a perturbation $W$. 
Perturbation theory assures us that $W$ contains only the dominant 
interactions, leaving out those that do not affect the qualitative
organization of the states. In addition having the ideal $H_0$ results in
 the
classical Hamiltonian being formulated from the start in action and angle
variables belonging to $H_0$. For a potential surface it is very 
complicated to construct an appropriate $H_0$ and the corresponding
action/angle variables. Our knowledge of the correct $H_0$ allows
the continuous deformation connection of some eigenstates of $H$ with the
corresponding eigenstates of $H_0$ which was useful for many states.
Second in the algebraic Hamiltonian the polyad ( which for the real 
molecule is only an approximate conserved quantity )
is an exactly conserved quantity and allows the reduction of the 
number of degrees of freedom. We always deal with one degree of freedom
less than the number of degrees of freedom on the potential surface.
Third, our strategy depends to a large extent on having a
compact configuration space with nontrivial homology where the
phase of the wave function is important and where the phase advances
divided by $2 \pi$ along the fundamental cycles of the configuration
space provide a large part of the excitation numbers. In part also 
feature 1 is responsible for the importance of the phases. This is an
essential difference to working in usual position space.
In total we have found a semiclassical representation of the states which
is actually much simpler to view than the usual position space one.

The classical dynamics of the system shows rather little regular motion
and is strongly unstable. Then the question arises in which way the
quantum states reflect anything from the classical chaos. We propose
the following answer:
In the middle of the polyad a large number of states do not show any
simple organization pattern. For some of those states it might be that
we simply have not found the right representation in which to see the
organizational center. However we doubt that this is the case for many
states. For many of the complicated states the wave function is really
without a simple structure, it seems to be a complicated interference
between several organizational structures. Then immediately comes the
idea to mix these states with their neighboring states to extract simple
structures in the way as we did it for example for states 22 and 24. 
For most of the
complicated states this method does not lead to any success. This
indicates that underlying simple patterns are distributed over many
states so that demixing becomes rather hopeless and maybe also meaningless. 
In the spirit of
reference \cite{bsw} this indicates that the density of states is already
so high that many states fall into the energy interval over which states
are considered almost degenerate and mix strongly. This is a strong
difference to the systems we have investigated before which could be
reduced to 2 degrees of freedom. There the density of states was much
lower such that mixing only occurred occasionally between 2 neighboring
states. This stronger mixing gives some clue as to how the quantum 
wave functions
become complicated for classically chaotic systems and for sufficiently high
quantum excitation.

Finally there may come up the question why we do not use some of the
common statistical procedures ( like e.g. random matrix theory ) to treat
our system when effects of classical chaos become evident in a quantum
system. The answer is: We are not satisfied with only generic
(universal) properties of the system or with the investigation to
which extent such universal properties are realized by our system.
We aim to investigate each individual quantum state one
after the other and to work out its individual properties in order to arrive
in the best case at a complete set of quantum numbers and in addition
at a detailed description of the atomic motion underlying the individual
state. Previous examples have shown that our procedure provides a
complete classification for simpler systems which can be reduced to
2 degrees of freedom. As the example presented in this paper shows, 
this program
is doable with good success for a large number of states even in a
system with originally 4 degrees of freedom, reducible 
only to 3 degrees of freedom and with a complicated classical dynamics.

\section*{Acknowledgments}
We thank Dr. E. Atilgan for helpful discussions.
CJ thanks CONACyT for support under grant number 33773-E and DGAPA for
support under grant number IN-109201.
HST thanks the US Department of Energy, Chemical Science Division for
support under grant number DE-FG03-01IR15147.

\clearpage

\begin{table}
\begin{center}
\caption{\label{tab1}
This table gives the classification and assignment of all states
organized around planes $\psi_f = \psi_b + constant$.
The first column gives the number of the state ordered according to energy.
If the structure only becomes clear after demixing, then it contains in
addition the number(s) of the state(s) with which it is demixed.
The second column gives the energy in units of $cm^{-1}$. The third and fourth
columns give the two longitudinal quantum numbers $l_a$ and $l_{b+f}$.
The fifth column gives the approximate symmetry of the state. The sixth
column gives the transverse quantum number $t$ relative to the planes
$\psi_f = \psi_b$ and $\psi_f = \psi_b + \pi$. The seventh column gives
the transverse quantum number $t'$ relative to the planes
$\psi_f = \psi_b \pm \pi/2$. If the transverse structure is washed out in
the figures then the corresponding transverse quantum numbers are put
in parenthesis. For such states where the individual
quantum numbers $n_b$ and $n_f$ exist, these are given in columns 8 and 9.
If these phase functions show some first indications of developing
singularities, these quantum numbers are put in brackets. States where
the pattern is not so clear in total but where the general trends suggest the
assignment have a $*$ in the last column.}

\begin{tabular}{c|c|c|c|c|c|c|c|c|c}
Number(s) & Energy & $l_a$ & $l_{b+f}$ & sym & $t$ & $t'$ &
$n_b$ & $n_f$ & \\ \hline
1        & 11564  & 0  & 10  & +  &  0  &     & 10  &  0  &    \\
2        & 11609  & 0  & 10  & -  &  0  &     &  9  &  1  &    \\
3        & 11724  & 1  &  9  & -  &  0  &     &  9  &  0  &    \\
4/5      & 11764  & 1  &  9  & +  &  0  &     &  8  &  1  &    \\
5/4      & 11775  & 0  & 10  & +  &  1  &     &  8  &  2  &    \\
6        & 11901  & 0  & 10  & -  &  1  &     &  7  &  3  &    \\
7        & 11917  & 2  &  8  & +  &  0  &     &  8  &  0  &    \\
8        & 11922  & 1  &  9  & -  &  1  &     &  7  &  2  &    \\
9        & 11964  & 2  &  8  & -  &  0  &     &  7  &  1  &    \\
10       & 12028  & 1  &  9  & +  &  1  &     & (6) & (3) &    \\
11       & 12060  & 0  & 10  & +  &  2  & (3) &  6  &  4  &    \\
12       & 12097  & 2  &  8  & +  &  1  &     & (6) & (2) &    \\
13       & 12123  & 3  &  7  & -  &  0  &     & (7) & (0) &    \\
14       & 12161  & 3  &  7  & +  &  0  &     &     &     &    \\

\end{tabular}
\end{center}
\end{table}

\newpage

\begin{table}
\begin{center}
\begin{tabular}{c|c|c|c|c|c|c|c|c|c}
Number(s) & Energy & $l_a$ & $l_{b+f}$ & sym & $t$ & $t'$ &
$n_b$ & $n_f$ &\\ \hline

15/16    & 12195  & 1  &  9  & -  & (2) & (2) & (5) & (4) &    \\
16/15/18 & 12225  & 2  &  8  & -  &  1  &  2  &     &     &    \\
17       & 12235  & 2  &  8  & +  &  2  &  2  &     &     &    \\
18/15/16 & 12240  & 0  & 10  & -  &  2  &  2  & (5) & (5) &    \\
19       & 12254  & 0  &  8  & +  &  0  &     &  8  &  0  &    \\
20       & 12283  & 1  &  9  & +  & (2) &  2  &     &     &    \\
21       & 12302  & 3  &  7  & -  &  1  & (2) & (5) & (2) &    \\
22/24    & 12318  & 4  &  6  & +  &  0  &     & (6) & (0) &    \\
23       & 12327  & 0  &  8  & -  &  0  &     &  7  &  1  &    \\
24/22    & 12358  & 0  & 10  & +  & (3) &  2  &     &     & *  \\
25       & 12380  & 4  &  6  & -  &  0  &     &     &     &    \\
26       & 12385  & 3  &  7  & +  &  1  &  2  &     &     & *  \\
27       & 12422  & 1  &  7  & -  &  0  &     &  7  &  0  &    \\
28/29    & 12466  & 4  &  6  & +  &  1  &  2  &     &     &    \\
29/28    & 12472  & 0  &  8  & +  &  1  &     &  6  &  2  &    \\
30       & 12495  & 1  &  7  & +  &  0  &     &  6  &  1  &    \\
31/34    & 12521  & 1  &  9  & -  &     &  1  &     &     & *  \\
32       & 12527  & 1  &  9  & +  &     &  1  &     &     &    \\
33       & 12539  & 5  &  5  & -  &  0  &     &     &     &    \\
34/31    & 12544  & 2  &  8  & -  & (2) &  1  &     &     & *  \\
35       & 12549  & 5  &  5  & +  &  0  &     &     &     &    \\
36       & 12561  & 2  &  8  & +  &     &  1  &     &     & *  \\
37       & 12575  & 0  & 10  & -  &     &  1  &     &     &    \\
38/39    & 12587  & 0  & 10  & +  &     &  1  &     &     &    \\
39/38    & 12590  & 3  &  7  & -  &     &  1  &     &     &    \\
40       & 12615  & 2  &  6  & +  &  0  &     &  6  &  0  &    \\
41/42    & 12622  & 3  &  7  & +  &     &  1  &     &     & *  \\
42/41    & 12622  & 0  &  8  & -  &  1  &     & (5) & (3) &    \\
43       & 12634  & 1  &  7  & -  &  1  & (2) &     &     &    \\
44       & 12635  & 4  &  6  & -  &  1  &  1  &     &     & *  \\
45       & 12647  & 4  &  6  & +  & (2) &  1  &     &     & *  \\
46       & 12689  & 2  &  6  & -  &  0  &     &  5  &  1  &    \\
47       & 12716  & 5  &  5  & -  &  1  &  1  &     &     &    \\
48/49/51 & 12739  & 5  &  5  & +  &  1  &  1  &     &     &    \\
49/48/51 & 12743  & 1  &  7  & +  &  1  &     & (4) & (3) &    \\
50       & 12745  & 6  &  4  & +  &  0  &     &     &     & *  \\
51/48/49 & 12762  & 0  &  8  & +  &  2  &  2  & (4) & (4) & *  \\
52       & 12793  & 6  &  4  & -  &  0  &     &     &     &    \\
53       & 12807  & 2  &  6  & +  &  1  &  2  &     &     & *  \\
54/55    & 12816  & 3  &  5  & -  &  0  &     &  5  &  0  &    \\
55/54    & 12822  & 0  &  6  & +  &  0  &     & (6) & (0) &    \\
56       & 12849  & 6  &  4  & +  &  1  &  1  &     &     & *  \\
57       & 12874  & 3  &  5  & +  &  0  &     &     &     &    \\
58       & 12906  & 1  &  7  & -  & (2) &  1  &     &     & *  \\

\end{tabular}
\end{center}
\end{table}

\newpage

\begin{table}
\begin{center}
\begin{tabular}{c|c|c|c|c|c|c|c|c|c}
Number(s) & Energy & $l_a$ & $l_{b+f}$ & sym & $t$ & $t'$ &
$n_b$ & $n_f$ &\\ \hline

59/61    & 12911  & 2  &  8  & -  &     &  0  &     &     & *  \\
60/63    & 12912  & 0  &  6  & -  &  0  &     & (5) & (1) &    \\
61/59    & 12916  & 7  &  3  & +  &  0  & (1) &     &     &    \\
62/64    & 12921  & 3  &  7  & -  &     &  0  &     &     &    \\
63/60    & 12929  & 0  &  8  & -  &  2  &  1  &     &     & *  \\
64/62    & 12933  & 2  &  8  & +  &     &  0  &     &     &    \\
65/66    & 12938  & 3  &  7  & +  &     &  0  &     &     &    \\
67       & 12943  & 1  &  9  & +  &     &  0  &     &     &    \\
68/69    & 12947  & 7  &  3  & -  &  0  &     & (3) & (0) &    \\
69/68    & 12951  & 1  &  9  & -  &     &  0  &     &     &    \\
70       & 12955  & 2  &  6  & -  & (1) & (1) &     &     & *  \\
73       & 12966  & 4  &  6  & +  &     & (0) &     &     & *  \\
74/78/81 & 12993  & 4  &  4  & +  &  0  &     & (4) & (0) & *  \\
76/77    & 13003  & 0  & 10  & -  &     &  0  &     &     &    \\
78/74    & 13006  & 0  & 10  & +  &     &  0  &     &     &    \\
79       & 13011  & 5  &  5  & +  &     &  0  &     &     & *  \\
81/74/78 & 13031  & 2  &  6  & +  &     &  1  &     &     & *  \\
84/85    & 13066  & 0  &  6  & +  &  1  &     & (4) & (2) & *  \\
85/84/86 & 13075  & 3  &  5  & +  &  1  &  1  &     &     & *  \\
86/84/85 & 13077  & 6  &  4  & +  & (2) &  0  &     &     & *  \\
88       & 13102  & 1  &  5  & +  &  0  &     &  4  &  1  &    \\
89       & 13125  & 7  &  3  & -  &  1  & (0) & (1) & (2) & *  \\
90       & 13132  & 7  &  3  & +  &  1  &  0  & (2) & (1) & *  \\
91       & 13152  & 4  &  4  & +  &  1  &  1  &     &     & *  \\
95/96    & 13206  & 0  &  6  & -  &  1  &  1  &  3  &  3  &    \\
97/99    & 13219  & 5  &  3  & +  &  0  &     &     &     & *  \\
100      & 13241  & 1  &  7  & -  &     &  0  &     &     & *  \\
101      & 13244  & 1  &  7  & +  &     &  0  &     &     &    \\

102      & 13248  & 2  &  6  & -  &     &  0  &     &     &    \\
103      & 13252  & 2  &  6  & +  &     &  0  &     &     & *  \\
104/105  & 13275  & 0  &  8  & -  &     &  0  &     &     & *  \\
105/104  & 13277  & 0  &  8  & +  &     &  0  &     &     & *  \\
107/108/109 & 13286 & 3 & 5  & -  &     &  0  &     &     & *  \\
108/109  & 13292  & 3  &  5  & +  &     &  0  &     &     &    \\
109/107/108 & 13298 & 2 & 4  & -  &  0  &     &     &     & *  \\
115/113  & 13333  & 1  &  5  & +  &  1  & (1) &     &     & *  \\
116/117  & 13359  & 4  &  4  & +  &     &  0  &     &     & *  \\
121      & 13418  & 0  &  4  & -  &  0  &     & (3) & (1) &    \\
123      & 13457  & 3  &  3  & +  &  0  &     &     &     &    \\
125/126/127 & 13494 & 1 & 3  & -  &  0  &     & (3) & (0) & *  \\
126/125/127 & 13497 & 0 & 6  & -  &     &  0  &     &     & *  \\
127/126  & 13501  & 1  &  5  & -  &     &  0  &     &     &    \\
128      & 13505  & 0  &  6  & +  &     &  0  &     &     & *  \\
139      & 13620  & 3  &  3  & +  &     & (1) &     &     &    \\

\end{tabular}
\end{center}
\end{table}

\begin{table}
\begin{center}
\caption{\label{tab2}
This table gives the classification and assignment of all states
organized around fibers in b direction.
The first column gives the number of the state ordered according to energy.
If the structure only becomes clear after demixing, then it contains in
addition the number(s) of the state(s) with which it is demixed.
The second column gives the energy in units of $cm^{-1}$. The third
column gives the longitudinal quantum number $l_b$.
The fourth and fifth
column give the transverse quantum numbers $t_1$ and $t_2$.}

\begin{tabular}{c|c|c|c|c}
Number(s) & Energy & $l_b$ & $t_1$ & $t_2$ \\ \hline
161       & 14172  &  0    &   0   &   0   \\
160       & 14074  &  0    &   1   &   0   \\
159/158   & 13983  &  1    &   0   &   0   \\
158/159   & 13980  &  0    &   2   &   0   \\
157       & 13929  &  0    &   0   &   1   \\
156/155   & 13893  &  1    &   1   &   0   \\
155/156   & 13887  &  0    &   3   &   0   \\
154       & 13853  &  0    &   1   &   1   \\
153/152   & 13814  &  1    &   2   &   0   \\
152/153   & 13805  &  0    &   2   &   1   \\
151       & 13782  &  1    &   0   &   1   \\

\end{tabular}
\end{center}
\end{table}

\clearpage

\begin{list}{}{\leftmargin 2cm \labelwidth 1.5cm \labelsep 0.5cm}

\item[\bf Fig. 1] 
Plots of density ( part a ) and phase ( part b ) 
respectively in the plane $\psi_f = \psi_b$ and of density ( part c ) 
and phase ( part d ) respectively in the plane $\psi_a = \pi $ for the 
state 7. In parts a and c darker grey means higher density. In parts b and d 
points with positive real part of the wave function are marked by a dot and 
points with a positive imaginary part of the wave function are marked by 
an open circle. All frame boundaries run from 0 to $2 \pi$.

\item[\bf Fig. 2] 
Plots for state 43 in the plane $\psi_f = \psi_b + \pi/8$
( parts a and b ) and in plane $\psi_f = \psi_a + \pi$ ( parts c and d ).
Otherwise this plot is done in analogy to Fig.1.

\item[\bf Fig. 3] 
Plots for state 160 in the plane $\psi_b = 0$
( parts a and b ) and in plane $\psi_f = - \psi_a$ ( parts c and d ).
Otherwise this plot is done in analogy to Fig.1.

\item[\bf Fig. 4]  
Plots for state 82 in the plane $\psi_f = 0$ ( parts a and b ) 
and in plane $ \psi_b = \psi_a + 0.29 * 2 \pi$ ( parts c and d ).
Otherwise this plot is done in analogy to Fig.1.

\item[\bf Fig. 5]
Plots for state 80 in the plane $\psi_b = \psi_a + \psi_f + \pi/2$
( parts a and b ) and in plane $\psi_f = 0$ ( parts c and d ).
Otherwise this plot is done in analogy to Fig.1.

\end{list}

\clearpage

\begin{figure}[ht]
\epsfig{file=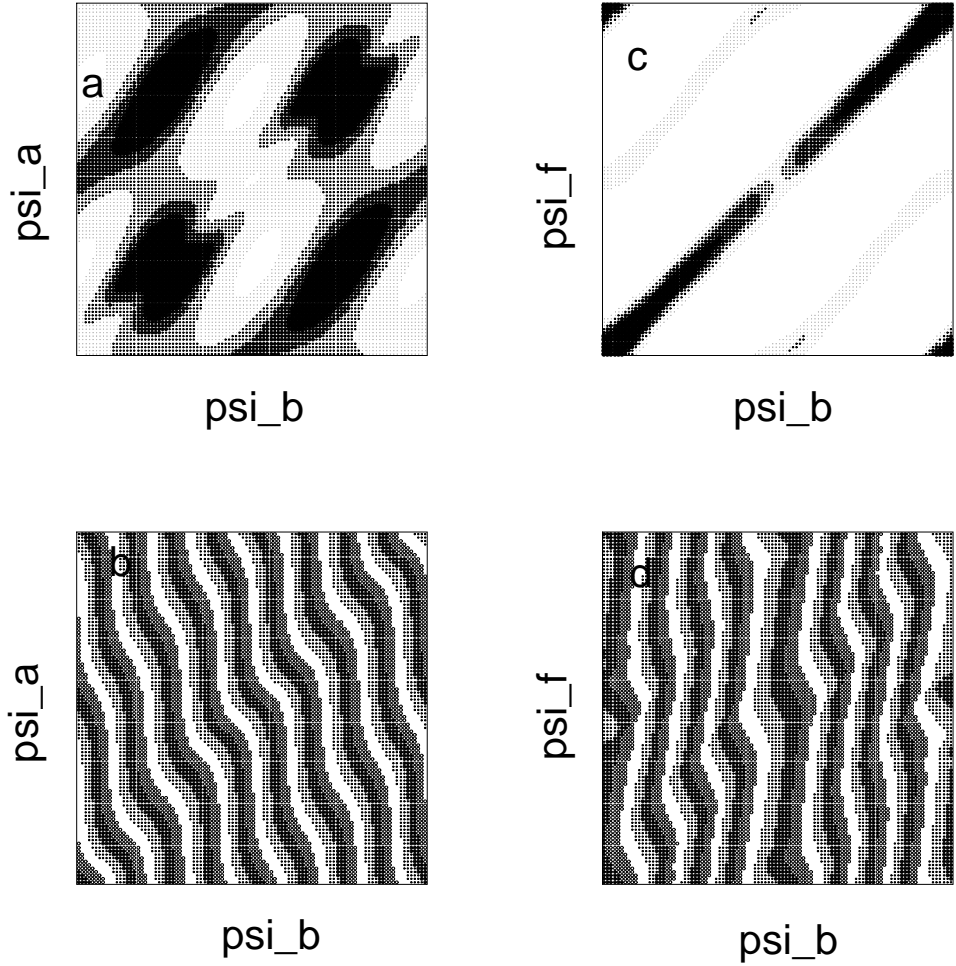, angle=-90, width=\columnwidth}
\caption{
Plots of density ( part a ) and phase ( part b ) respectively in
the plane $\psi_f = \psi_b$ and of density ( part c ) and phase ( part d )
respectively in the plane $\psi_a = \pi $ for the state 7. In parts a and c
darker grey means higher density. In parts b and d points with positive
real part of the wave function are marked by a dot and points with a
positive imaginary part of the wave function are marked by an open circle.
All frame boundaries run from 0 to $2 \pi$.}
\end{figure}

\begin{figure}[ht]
\epsfig{file=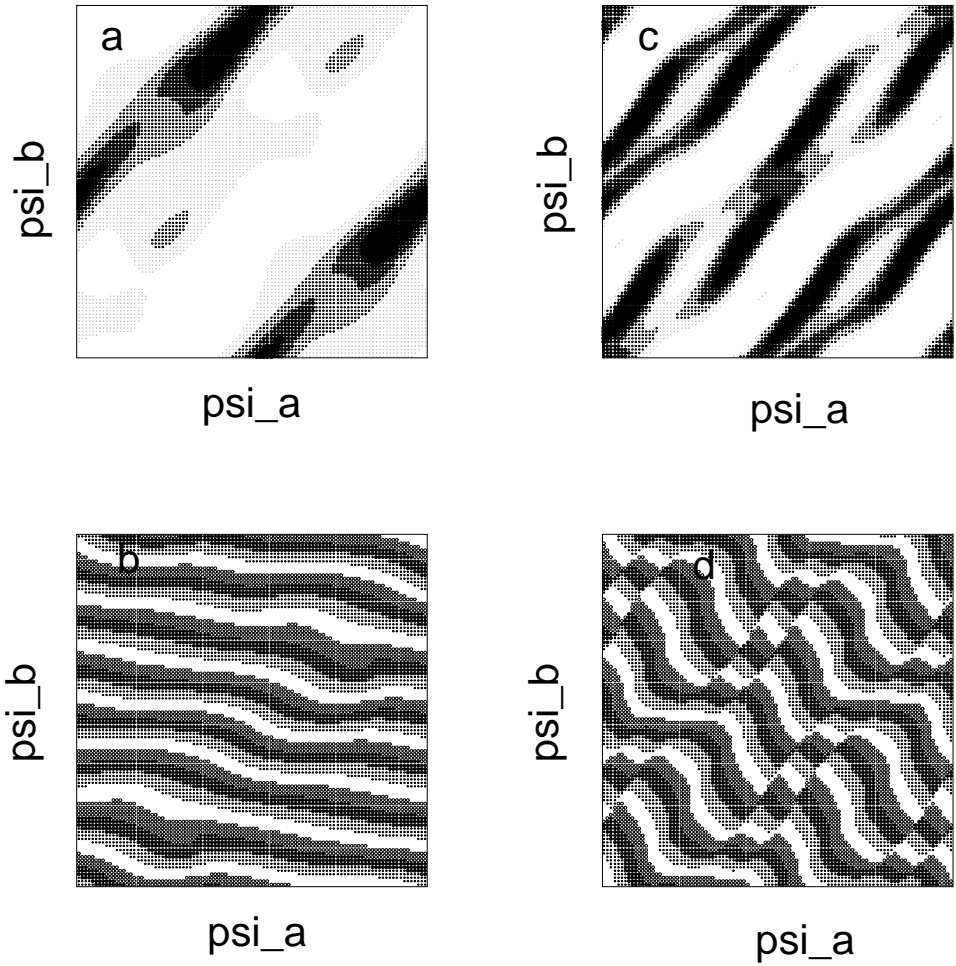, angle=-90, width=\columnwidth}
\caption{
Plots for state 43 in the plane $\psi_f = \psi_b + \pi/8$
( parts a and b ) and in plane $\psi_f = \psi_a + \pi$ ( parts c and d ).
Otherwise this plot is done in analogy to Fig.1.
}
\end{figure}

\begin{figure}[ht]
\epsfig{file=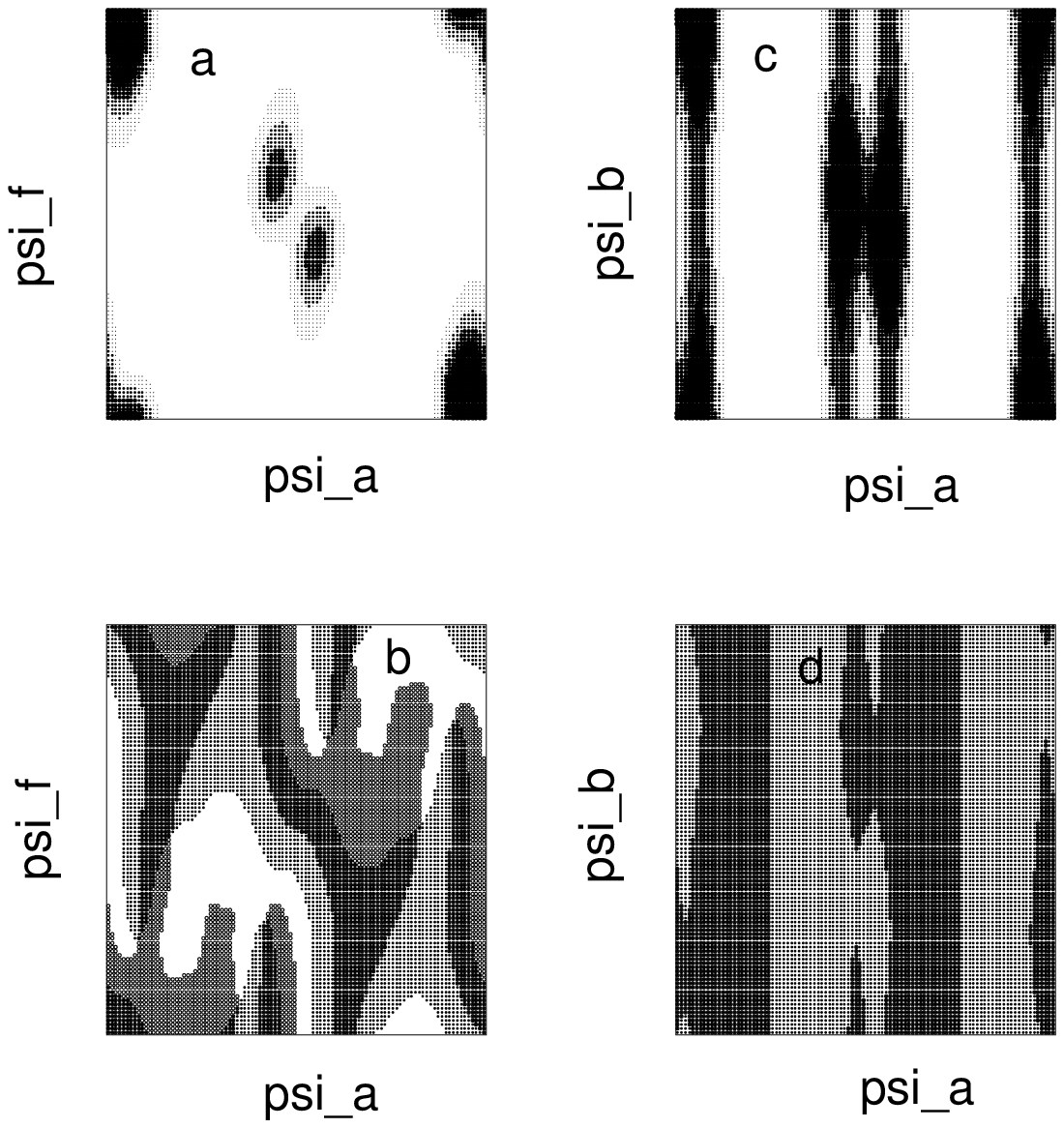, angle=-90, width=\columnwidth}
\caption{
Fig.3: Plots for state 160 in the plane $\psi_b = 0$
( parts a and b ) and in plane $\psi_f = - \psi_a$ ( parts c and d ).
Otherwise this plot is done in analogy to Fig.1.
}
\end{figure}

\begin{figure}[ht]
\epsfig{file=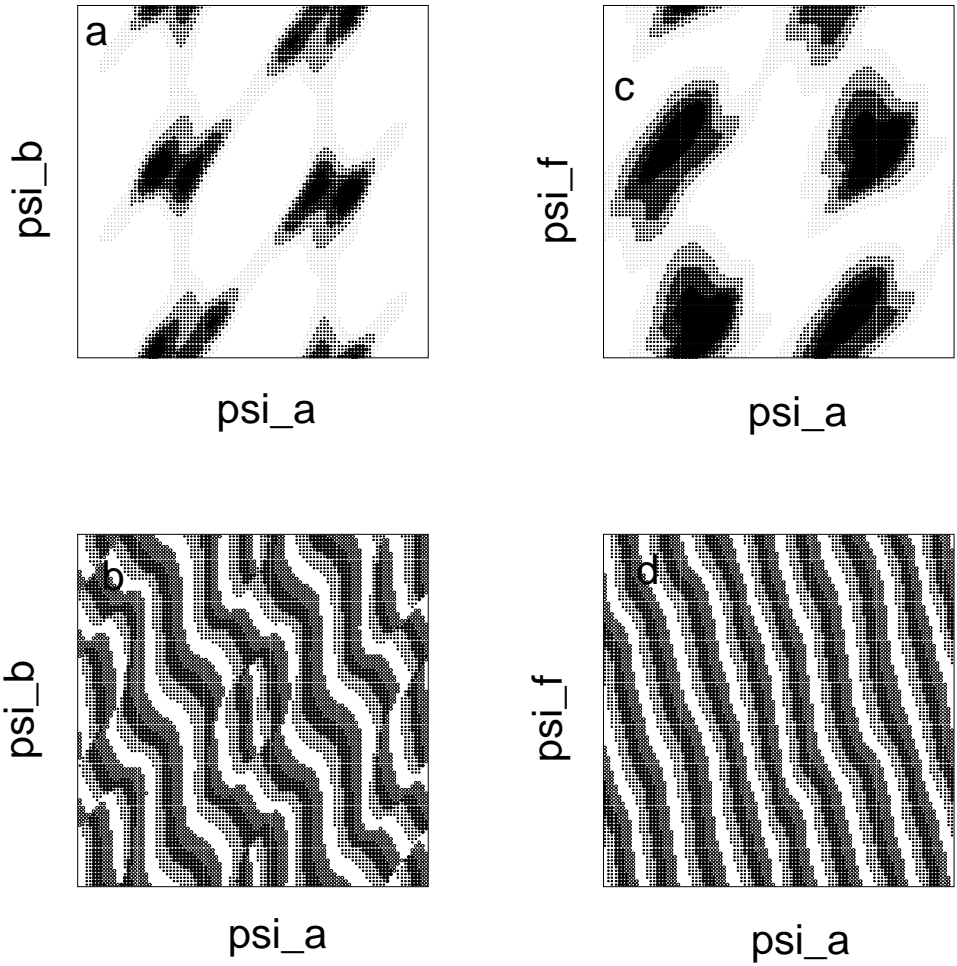, angle=-90, width=\columnwidth}
\caption{
Plots for state 82 in the plane $\psi_f = 0$ ( parts a and b ) and in plane
$ \psi_b = \psi_a + 0.29 * 2 \pi$ ( parts c and d ).
Otherwise this plot is done in analogy to Fig.1.
}
\end{figure}

\begin{figure}[ht]
\epsfig{file=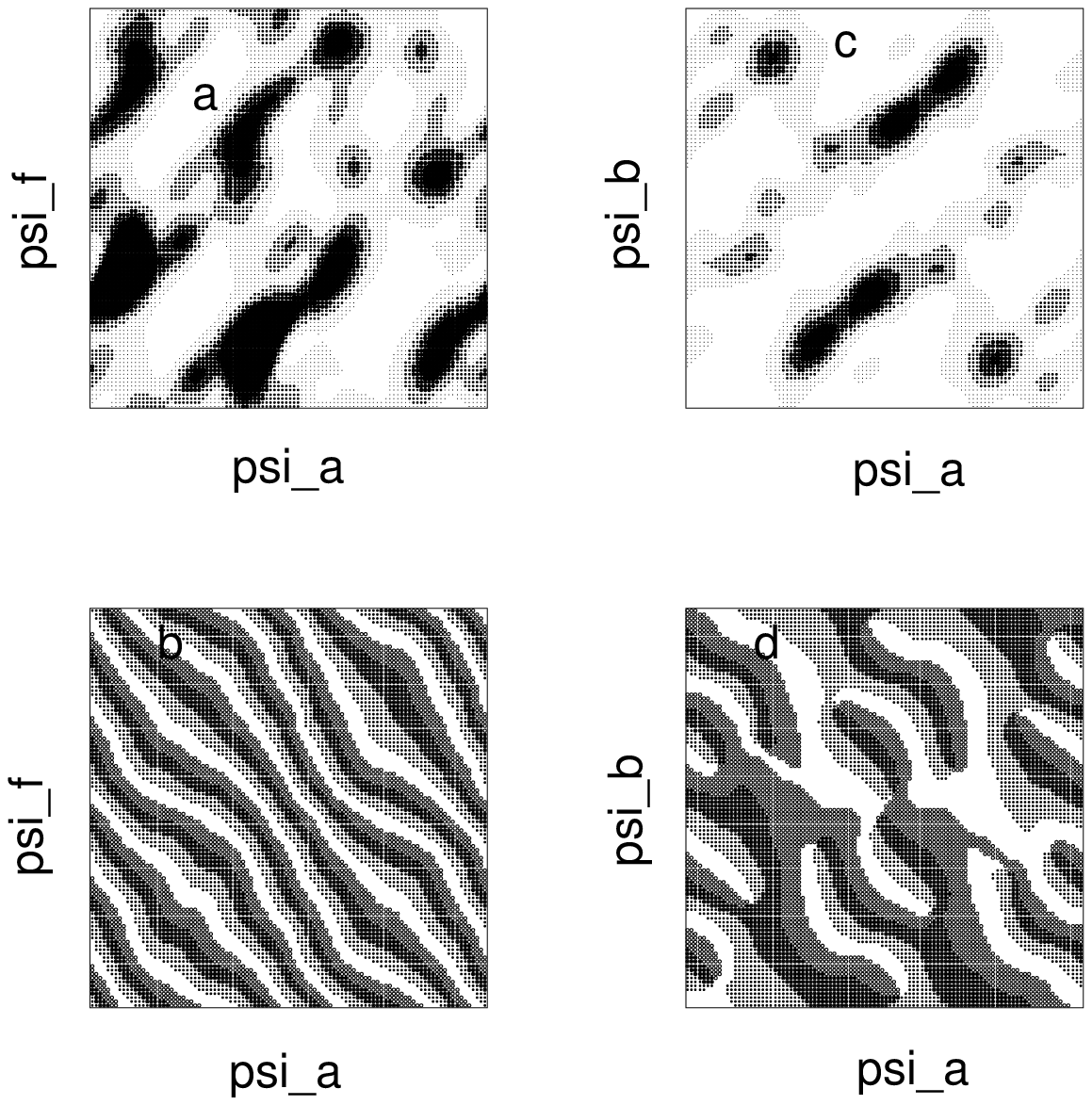, angle=-90, width=\columnwidth}
\caption{
Fig.5: Plots for state 80 in the plane $\psi_b = \psi_a + \psi_f + \pi/2$
( parts a and b ) and in plane $\psi_f = 0$ ( parts c and d ).
Otherwise this plot is done in analogy to Fig.1.
}
\end{figure}

\end{document}